\documentclass[aps,preprint,superscriptaddress,floatfix]{revtex4}%
\usepackage{amsfonts}
\usepackage{amsmath}
\usepackage{amssymb}
\usepackage{graphicx}
\usepackage{graphicx}%
\begin{document}
\title{Powder vibration studied by M\"{o}ssbauer spectroscopy}
\author{A. L. Zinnatullin}
\affiliation{Kazan Federal University, 18 Kremlyovskaya Street, Kazan 420008 Russia}
\author{R. N. Shakhmuratov}
\affiliation{Kazan Federal University, 18 Kremlyovskaya Street, Kazan 420008 Russia}
\affiliation{Zavoisky Physical-Technical Institute, FRC Kazan Scientific Center of RAS,
Kazan 420029, Russia}
\author{F. G. Vagizov}
\affiliation{Kazan Federal University, 18 Kremlyovskaya Street, Kazan 420008 Russia}
\date{{ \today}}

\begin{abstract}
M\"{o}ssbauer spectroscopy is applied to study ultrasound vibration of a
granular material. A pile of powder is placed onto the surface of piezo
transducer vibrated with the frequency 12.68 MHz. The size of grains (1.3
$\mu$m) is much smaller than the wavelength of ultrasound, but much larger
than the vibration amplitude. Due to vibration a single line of the
M\"{o}ssbauer transmission spectrum is split into a comb structure with a
period equal to the vibration frequency. This spectrum contains the
information about fast and slow modes in granular dynamics. We developed a
method which allows to measure decay of ultrasound in the granular material
and to estimate the results of the particle convection of grains in the pile.

\end{abstract}
\maketitle

\section{Introduction}

A granular material is conventionally considered as a conglomeration of
discrete macroscopic objects, which are larger than that composing mesoscopic
or microscopic media. The size of particles making up a granular medium is
large enough such that they are not subject to thermal Brownian motion.
Therefore, the lower limit of particle diameter is about 1 $\mu$m
\cite{Duran}. Collective behavior of a granular medium is characterized by a
loss of energy, for example, due to friction when particles collide. However,
despite its simplicity such a system behaves differently from any of the other
familiar forms of matter - solids, liquids, or gases. If granular materials
are excited, for example, vibrated or allowed to flow, grains exhibit a wide
range of pattern forming not a single phase of matter but have characteristics
reminiscent of solids, liquids, or gases \cite{Jaeger}. The famous Brazil nuts
effect, demonstrating that the larger or lighter grains are always found on
top of the shaken box of multigrain m\"{u}esli, is explained via a model that
includes sequential as well as nonsequential (cooperative) particle dynamics
\cite{Ristow}. Avalanche of particles, bulk convection of grains to the top of
the pile, phase transition are phenomena that demonstrate exceptional
properties of granular materials
\cite{Ristow,Eshuis,Herrmann,Klongboonjit,Liu,Nagel}. They play an important
role in industries, such as mining, agriculture, construction, pharmaceutics,
etc. The extraction of ores, sands, and gravel, which relies on dredging,
crushing and grinding, followed by separation, are commonly used for
production of granular materials. Therefore, improvement and optimization in
methods of transport, storage and mixing would have a major economic impact.
The cosmetic and pharmaceutical industries, specialized chemistry, and the
food industry demand increasingly sophisticated processing technologies when
it is absolutely essential to achieve intimate mixing of different granular
materials without their separation. Ultrasonic microfeeding of fine powders is
a promising method for solid freeform fabrication by 2D and 3D printing and
pharmaceutical dosing \cite{Lu}.

Recent scientific progress in knowledge about open nonequilibrium systems and
development of modern computational methods give impetus to studying of
granular dynamics \cite{Ristow}. Granular material is a multiparticle system.
Therefore, many mathematical models are simplified to one-dimensional or
two-dimensional examples. Interparticle forces in granular materials such as
the static friction at particle contacts, their collision, and particle-wall
contacts are well studied for two-dimensional systems \cite{Majmudar}.
Analysis of real-life three-dimensional situations is still quite complicated.
However, with some assumptions and simplifications the problem can be solved
for model systems \cite{Bougie,Harada,Amirifar}.

Experimentally, the low frequency vibration of granular material was studied
by optical methods in two-dimensional system (8 to 30 Hz vibration)
\cite{Clement}, by means of nuclear magnetic resonance (NMR) of poppy seeds in
three-dimensional system subject to the periodic sequence of shakes (each
shake consists of a single 20-Hz sinusoidal period of acceleration separated
by 0.7 s between shakes) \cite{Nagel2}. Measurements in three-dimensional
medium, vibrated with frequency 30-100 Hz, was also performed by observation
of a thin radioactive-marker layer sandwiched between studied granular layers
\cite{Harwood}. In this experiment migration of marker could be followed in
either an upward or downward direction.

High frequency vibration over frequency range from 100 to 350 kHz is used in
experiments studying marine sediments, which are composed of unconsolidated
granular materials \cite{Buck}. Gaussian burst signals with center frequencies
0.4 - 1.6 MHz where used to excite ultrasound pulses in granular media by
piezoelectric transducer \cite{Zhai}. Piezoelectric transducer also were
employed in this experiment to detect longitudinal sound waves and X-ray
diffraction provided tomography of contact fabric, particle kinematics,
average per-particle stress tensors, and interparticle forces.

Ultrasound vibration of granular medium induces slow and fast modes in
granular dynamics. Time scales of these modes are well separated \cite{Mehta}.
Transport of grains producing bulk convection, size segregation, and dynamical
phase transition are relatively slow. They can be studied by optical methods,
X-ray diffraction, or NMR tomography. Fast modes are difficult to trace. Only
attenuation of the longitudinal sound wave can be detected at the exit of the
granular medium excited from the opposite side \cite{Zhai}.

It is interesting to note that, for example, a column of beads or other
elastic granular material vibrated vertically demonstrate distinct behavior
depending on the acceleration $A_{\text{acs}}=a\Omega^{2}$ given by the
vibrated plate at the bottom of the column, where $a$ and $\Omega$ are the
amplitude and frequency of the plate vibration. When acceleration is small,
the system is in a "condensed" state where the beads are practically in
contact with each other moving in unison. When acceleration is large, the
system is in a "fluidized" state, with the beads moving individually much like
particles in a gas or a fluid \cite{Duran}. In this state the kinetic energy
$E_{\text{kin}}=ma^{2}\Omega^{2}/2$ is the main parameter defining the system
dynamics. Transition between two states is defined by a parameter
$\Lambda=A_{\text{acs}}/g$, where $g$ is gravitational acceleration. Usually,
when $\Lambda>1$ the qualitative behavior of particles changes. However, this
condition depends on individual particle mass and friction force \cite{Duran}.

In this paper we study ultrasonic vibration of powder consisting of grains of
potassium ferrocyanide K$_{4}$Fe(CN)$_{6}\cdot$3H$_{2}$O. It is the potassium
salt, which forms monoclinic crystal. The salt is grind up to produce
crystallites with sizes satisfying the lognormal distribution with a median
size of 1.3 $\mu$m and a standard deviation of 0.18. This powder is vibrated
by polymer (PVDF) piezoelectic transducer with frequency 12.68 MHz. The
wavelength of the ultrasound generated in the solid grains is several hundred
microns, i.e., much larger than the size of an individual particle. The
amplitude $a$ of ultrasound vibration, induced by piezoelectic transducers, is
usually extremely small. Our previous experiments showed that thin PVDF
transducer is capable producing 25 $\mu$m thick stainless-steel foil vibration
with frequency 12.68 MHz and amplitude $a$ about one angstrom
\cite{Shakhmuratov1,Shakhmuratov2}. Rough estimation of the acceleration
$A_{\text{acs}}$ imparted to the objects placed on the surface of the
transducer gives the value $5.4\times10^{5}$ m/s$^{2}$, which is comparable
with a bullet's acceleration in the barrel of a pistol. Therefore the
parameter $\Lambda$ is extremely large.

Vibration of the salt grains, enriched by $^{57}$Fe isotope, is studied by
M\"{o}ssbauer spectroscopy where 14.4-keV $\gamma$-radiation is used to
observe the absorption spectrum of $^{57}$Fe nuclei. This method has many
advantages since the wavelength of such a radiation is slightly below one
angstrom (86 pm) and quality factor $Q$ of the $^{57}$Fe absorption line is
extremely large, i.e., $Q=\omega_{A}/\Gamma_{A}=3\times10^{12}$, where
$\omega_{A}$ is the resonant frequency and $\Gamma_{A}$ is the width of the
absorption line.

When the quasi-monochromatic source radiation with main frequency $\omega_{S}$
is transmitted through a medium with resonant nuclei, which are vibrated, the
spectrum of the radiation field is transformed into polychromatic with a set
of spectral lines $\omega_{S}\pm n\Omega$, where $\Omega$ is the vibration
frequency and $n=0,\pm1,\pm2$ \cite{Shakhmuratov1,Shakhmuratov2,Cranshaw,
Mishroy,Asher,Tsankov,Shvydko,Shakhmuratov3}. These lines are produced by
Raman scattering, which is inelastic scattering of $\gamma$-radiation by
vibrating nuclei. The intensities of the spectral components and their
dependence on the vibration amplitude give the information about the decay of
ultrasound and the distribution of the vibration amplitudes along the medium.
Thus, our experiments are capable to provide the information about fast mode
in granular dynamics (ultrasound vibration and decay along the granular
medium) and verify the models of slow mode including slow flow and convection.

\section{Method}

In M\"{o}ssbauer spectroscopy a radiation source and absorber are basic
elements. For example, for the solid absorber containing $^{57}$Fe nuclei the
solid source matrix with inclusions of radioactive nuclei $^{57}$Co is usually
used. These nuclei decay randomly emitting $\gamma$-photons with the energy
14.4 keV and spectral width $\Gamma_{S}$. For a single line source the minimum
value of this width is $\Gamma_{0}=1.1$ MHz. Activity of the commercially
available sources allows to produce a random flow of single $\gamma$-photons
with the average rate about $10^{4}$ photons per second. Slow motion of the
radiation source with respect to the absorber gives the oppotunity to scan the
absorption line of $^{57}$Fe due to Doppler effect. To accumulate absorption
spectrum, the electric pulses produced by $\gamma$-detection system must be
recorded synchronously to the sweep velocity of the source. Collecting number
of counts corresponding to different velocities within fixed time windows, we
obtain absorption spectrum, which is seen as a drop of counts at resonance and
constant rate of counts far from resonance. Since the coherence length of
$\gamma$-photons is 141 ns, which is the duration of a single-photon pulse, we
have a snapshot of fast processes, for example, vibration of nuclei with
frequency larger than $\Gamma_{S}$. Typical data-acquisition time for one
M\"{o}ssbauer spectrum may range from several hours to several days, depending
on resonant nuclei concentration and absorber thickness. Therefore, these
spectra contain information about slow steady-state processes averaged over a
long time period and reflecting slow dynamics of grains with $^{57}$Fe nuclei
such as slow flow and convection.

The Fourier transform of the radiation field amplitude emitted by the source
nucleus is%
\begin{equation}
E_{S}(\omega)=\frac{E_{0}e^{ikz}}{\Gamma_{S}/2+i(\omega_{S}-\omega)}.
\label{Eq1}%
\end{equation}
where $E_{0}$ and $k$ are the amplitude and wave number of the radiation field
emitted by the source nucleus, respectively (see Appendix A).

Transmission probability of a single $\gamma$-photon through the resonant
absorber is described by equation (see Ref. \cite{Gutlich} and Appendix A)
\begin{equation}
N_{\text{out}}(\omega_{A}-\omega_{S})=\frac{\Gamma_{0}}{2\pi}\int_{-\infty
}^{\infty}\frac{\exp\left[  -\frac{T_{A}(\Gamma_{0}/2)^{2}}{\left(  \Gamma
_{A}/2\right)  ^{2}+(\omega_{A}-\omega)^{2}}\right]  }{\left(  \Gamma
_{S}/2\right)  ^{2}+(\omega_{S}-\omega)^{2}}d\omega, \label{Eq2}%
\end{equation}
where $\omega_{A}$ and $\Gamma_{A}$ are the frequency and linewidth of the
absorber nuclei, $T_{A}=f_{A}n_{A}\sigma_{A}$ is the effective (optical)
thickness of the absorber, $f_{A}$ is the Debye-Waller factor for nuclei in
the solid state absorber denoting the fraction of $\gamma$-emission or
absorption occurring without recoil, $n_{A}$ is the number of $^{57}$Fe nuclei
per unit area of the absorber, and $\sigma_{A}$ is the resonant absorption
cross section. Here, for simplicity, nonresonant absorption and the fraction
of the radiation field with recoil are disregarded. They can be easily taken
into account in experimental data analysis. Example of $N_{\text{out}}%
(\omega_{A}-\omega_{S})$ function is shown in Fig. 1 by dotted blue line.
\begin{figure}[ptb]
\resizebox{0.6\textwidth}{!}{\includegraphics{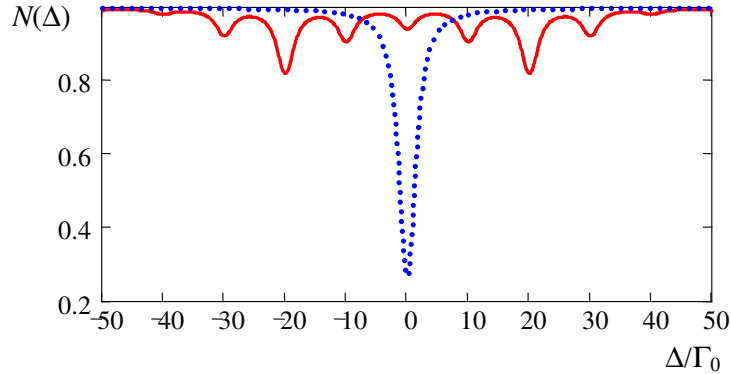}}\caption{(color on
line) Dependence of the detection probability on $\Delta=\omega_{A}-\omega
_{S}$. Blue dots correspond to $N_{\text{out}}(\Delta)$, which describes
standard M\"{o}ssbauer spectrum, Eq. (\ref{Eq2}). Red line shows M\"{o}ssbauer
spectrum, $N_{SV\text{out}}(\Delta)$, for the source vibrated with frequency
$\Omega=10\Gamma_{0}$ and modulation index $M=3$, see Eq. (\ref{Eq7}).
Effective thickness of the absorber is $T_{A}=5$}%
\label{fig:1}%
\end{figure}

\subsection{Vibrating source}

To explain the main points of the method used in this paper we will first
consider an example when the source nucleus vibrates with the frequency
$\Omega$ and the absorber is motionless. Then, the distance between the source
and absorber oscillates as $z_{\text{vib}}=z+a\sin(\Omega t+\psi)$, where $a$
and $\psi$ are the amplitude and phase of the vibration. The radiation field
of the source is $E_{S}(t-t_{0})=E_{0}\theta(t-t_{0})e^{-(i\omega_{S}%
+\Gamma_{0}/2)(t-t_{0})+ikz}$, where $t_{0}$ is the instant of time when the
excited state nucleus is formed in the source and $\theta(t-t_{0})$ is the
Heaviside step-function. Nuclei of the absorber interact with the field whose
phase is modulated due to oscillation of the distance $z_{\text{vib}}$.
Therefore, the source field modifies as follows%
\begin{equation}
E_{SV}(\tau)=E_{S}(\tau)e^{ika\sin[\Omega(\tau+t_{0})+\psi]}, \label{Eq3}%
\end{equation}
where $\tau=t-t_{0}$. According to the Jacobi-Anger formula this expression
can be presented as%
\begin{equation}
E_{SV}(\tau)=E_{S}(\tau)\sum_{n=-\infty}^{+\infty}J_{n}(M)e^{in[\Omega
(\tau+t_{0})+\psi)]}, \label{Eq4}%
\end{equation}
where $J_{n}(M)$ is the $n$th order Bessel function, $M=ka=2\pi a/\lambda$ is
the modulation index of the field phase, and $\lambda$ is the radiation
wavelength. Due to vibration of the source the single line radiation field
with frequency $\omega_{S}$ transforms into polychromatic, consisting of a set
of spectral lines $\omega_{S}-n\Omega$ with $n=0,\pm1,\pm2,...$ Fourier
transform of this field is%
\begin{equation}
E_{SV}(\omega)=E_{0}\sum_{n=-\infty}^{+\infty}\frac{J_{n}(M)e^{in\phi}}%
{\Gamma_{0}/2+i(\omega_{S}-n\Omega-\omega)}, \label{Eq5}%
\end{equation}
where $\phi=\Omega t_{0}+\psi$. Here and below for shortening $ikz$ is omitted
in the exponent. After passing through the absorber with a single resonance
line this field is transformed as%
\begin{equation}
E_{SV\text{out}}(\omega)=E_{0}\sum_{n=-\infty}^{+\infty}\frac{J_{n}%
(M)e^{-\frac{T_{A}\Gamma_{A}/4}{\Gamma_{A}/2+i(\omega_{A}-\omega)}+in\phi}%
}{\Gamma_{0}/2+i(\omega_{S}-n\Omega-\omega)}. \label{Eq6}%
\end{equation}
Calculating the probability of photon detection at the exit of the absorber in
the same way as in the case of the nonvibrating source (see Appendix A), we
obtain%
\begin{equation}
N_{SV\text{out}}(\Delta)=\sum_{n=-\infty}^{+\infty}J_{n}^{2}(M)L(\Delta_{n}),
\label{Eq7}%
\end{equation}
where $\Delta=\omega_{A}-\omega_{S}$, $\Delta_{n}=\omega_{A}-\omega
_{S}+n\Omega$, and
\begin{equation}
L(\Delta_{n})=\frac{\Gamma_{S}}{2\pi}\int_{-\infty}^{+\infty}\frac
{e^{-\frac{T_{A}(\Gamma_{0}/2)^{2}}{(\Gamma_{0}/2)^{2}+(\omega_{A}-\omega
)^{2}}}}{(\Gamma_{0}/2)^{2}+(\omega_{S}-n\Omega-\omega)^{2}}d\omega.
\label{Eq8}%
\end{equation}
This expression is valid if $\Omega\gg\Gamma_{0}$, which means that the
distance between the spectral lines of the source is much larger than the
linewidth of the absorber. Therefore, the contribution of cross terms, such as
$J_{n}(M)J_{m}(M)$, is negligible. Comparison of $N_{\text{out}}(\Delta)$ with
$N_{SV\text{out}}(\Delta)$ is shown in Fig. 1.

Far form resonance for all individual spectral components of the source with
the single line absorber, i.e., when $\left\vert \Delta_{n}\right\vert
/\Gamma_{0}\gg1$, the function $L(\Delta_{n})$ equals unity for all $n$. In
this case all the spectral components of the vibrating source pass through the
motionless absorber without interaction. Then, due to identity%
\begin{equation}
\sum_{n=-\infty}^{+\infty}J_{n}^{2}(M)\equiv1, \label{Eq9}%
\end{equation}
we have $N_{SV\text{out}}(\Delta)=1$.

When, for example, $n$th component comes close to resonance with the absorber
($\omega_{S}-n\Omega\approx\omega_{A}$), then%
\begin{equation}
N_{SV\text{out}}(\omega_{A}-\omega_{S})=1-J_{n}^{2}(M)\left[  1-L(\Delta
_{n})\right]  , \label{Eq10}%
\end{equation}
i.e., only resonant component of the field interacts with the absorber, while
other components pass through without interaction. This expression is derived
with the help of identity (\ref{Eq9}). At $\omega_{A}-\omega_{S}=-n\Omega$,
the detection probability of a single photon drops down to%
\begin{equation}
N_{SV\text{out}}(n\Omega)=1-J_{n}^{2}(M)\left[  1-N_{\text{out}}(0)\right]  .
\label{Eq11}%
\end{equation}
The scale of the decrease in the photon probability at the exit of the
absorber is defined by the square of the corresponding Bessel function,
$J_{n}^{2}(M)$.

\subsection{Vibrating absorber}

If the absorber vibrates with respect to the motionless source, the radiation
field $E_{AV}(\tau)$ in the coordinate system rigidly bounded to the absorber
sample is again acquires a comb structure with spectral componets $\omega
_{S}-n\Omega$ where $n=0,\pm1,\pm2,...$, i. e., $E_{AV}(\tau)=E_{SV}(\tau)$,
see Eq. (\ref{Eq4}).

We calculated the field, transmitted through the absorber, and transformed the
output field back to the lab frame (see Appendix A). If all spectral
components of the comb $E_{AV}(\tau)$ in the vibrating frame are far from
resonance with the single line absorber, then their amplitudes do not change.
In this case Fourier transform of the output field in the lab frame,
$E_{L}(\omega)T,$ does not change also, i.e., $E_{L}(\omega)=E_{S}(\omega)$.
Then, the probability of its registration is unity, i.e., equals to $N_{0}=1$.

If $n$th component of the comb is close to resonance with the absorber and
other spectral components are far from resonance, then%
\begin{equation}
E_{L}(\omega)=E_{S}(\omega)+E_{R}(\omega), \label{Eq12}%
\end{equation}%
\begin{equation}
E_{R}(\omega)=-E_{0}J_{n}(M)\sum_{m=-\infty}^{+\infty}\frac{J_{m}%
(M)e^{i(n-m)\phi}\left[  1-e^{-\frac{T_{A}\Gamma_{A}/4}{\Gamma_{A}%
/2+i(\omega_{A}+m\Omega-\omega)}}\right]  }{\Gamma_{0}/2+i[\omega
_{S}-(n-m)\Omega-\omega]}, \label{Eq13}%
\end{equation}
where $E_{R}(\omega)$ is the field scattered by the vibrating absorber. The
scattered field is polychromatic containing the resonant for the absorber
component with frequency $\omega_{S}=\omega_{A}+n\Omega$ and Raman components
$\omega_{S}-(n-m)\Omega=\omega_{A}+m\Omega$ with $m\neq n$. The resonantly
scattered component is in antiphase with the incident radiation reducing its
intensity due to destructive interference. The Raman components appear due to
inelastic scattering of the incident radiation field on the vibrating nuclei.
The amplitudes of these components depend on $J_{n}(M)$, i.e., on the
amplitude of the resonant component in the frequency comb $E_{AV}(\tau)$.

At exact resonance ($\omega_{S}=\omega_{A}+n\Omega$), Eq. (\ref{Eq13}) can be
expressed as%

\begin{equation}
E_{R}(\omega)=-E_{0}\left[  J_{n}^{2}(M)K_{n}(\omega)+J_{n}(M)\sum
_{\substack{m=-\infty\\(m\neq n)}}^{+\infty}J_{m}(M)K_{m}(\omega
)e^{i(n-m)\phi}\right]  , \label{Eq14}%
\end{equation}
where%
\begin{equation}
K_{m}(\omega)=\frac{\left[  1-e^{-\frac{T_{A}\Gamma_{A}/4}{\Gamma
_{A}/2+i(\omega_{A}+m\Omega-\omega)}}\right]  }{\Gamma_{0}/2+i(\omega
_{A}+m\Omega-\omega)}, \label{Eq15}%
\end{equation}
The maxima of the functions $K_{m}(\omega)$, which take place at frequencies
$\omega_{m}=\omega_{A}+m\Omega$, are the same for all spectral components $m$,
i.e., resonant $n$ and Raman components $m\neq n$. Therefore, the maximum
amplitudes of the spectral components are proportional to $J_{n}(M)J_{m}(M)$
with the same factor $K_{m}(\omega_{m})=2\left(  1-e^{-T_{A}/2}\right)
/\Gamma_{0}$.

In spite of the difference between the output fields for the vibrating source
and motionless absorber, Eq. (\ref{Eq6}), and for the source and vibrating
absorber, Eq. (\ref{Eq12}), the probability of photon detection at the exit of
the vibrated absorber is described by the same expression as for the vibrating
source, see Eqs. (\ref{Eq7}) and (\ref{Eq10}). This is because the
transformation back to the lab frame does not influence the time integrated
intensity of the radiation field, which coincides in the vibrating frame with
the radiation field of the vibrated source transmitted through the motionless absorber.

\section{The Model}

We studied M\"{o}ssbauer spectra of potassium ferrocyanide powder placed on
PVDF piezo transducer. The pile of powder was made thin (about 100 $\mu$m
thick) and perfectly flat, as well as possible, by stirring. Vibration of
grains with frequency $\Omega=12.68$ MHz splits the single absorption line
into a comb structure with a frequency period equal $\Omega$. This splitting
appears only after a long time preparation period. We suppose that the
preparation period is needed to form densely packed grains corresponding to
the solid phase located in the pile bottom on the surface of the transducer.
Otherwise, acoustic contact between piezo transducer and powder is not formed
and absorption line is not split. We noted that slow convection also took
place since small traces of grains were found around the pile after
preparation period and experimental data collection. Therefore, we suppose
that grains at the top of the pile form a gas-like phase and some of the
grains are thrown out from the pile top due to vibration.

Wavelength of ultrasound in potassium ferrocyanide is nearly four times large
than the thickness of the powder pile. Therefore, one could expect piston-like
vibration of the absorber. Example of the M\"{o}ssbauer transmission spectrum
for the grains vibrated in unison with the same amplitude is shown in Fig. 1.
Since the depth of $n$-th component of the spectrum is proportional to the
oscillating function $J_{n}^{2}(M)$, the pattern of the comb minima forms a
nonmonotonous envelope with oscillations. Similar spectra were observed in
experiments \cite{Shakhmuratov1,Shakhmuratov2,Mkrtchyan77,Mkrtchyan79} where
the vibration of the stainless steel (SS) foil was studied. It is also
possible even suppressing almost to nonabsorbing level the central component
of M\"{o}ssbauer spectrum with $n=0$ when modulation index is $M=2.4$ since
$J_{0}^{2}(2.4)=0$. This results in acoustically induced transparency of the
absorber for $\gamma$-radiation. Experimental implementation of the
transparency effect was demonstrated for SS foil in Ref. \cite{Radeon}.

Meanwhile, our M\"{o}ssbauer spectra for the vibrated powder demonstrate
qualitatively different pattern. Comb minima follow a bell-shape envelope and
the central component is always dipper than others for all values of the
modulation index, which depends on RF power applied to piezo transducer, see
Fig. (2). Similar spectra were observed for powder pressed in a tablet onto
the surface of piezo transducer \cite{Shakhmuratov1} and powder suspended in
viscous liquid or resin \cite{Cranshaw,Shakhmuratov3}. One may assume that
grains in a pile or grains in a resin vibrate with the same frequency but
their amplitudes are different depending on the distance from the transducer.
The phase of vibrations cannot be chaotic and randomly changing in a time
scale comparable with the vibration period $2\pi/\Omega$. Otherwise, the
components of the comb spectrum will be broadened progressively with increase
of the number $n$. This is because of the phase factor $e^{in\psi}$ in the
expression for the field $E_{AV}(\tau)=E_{SV}(\tau)$, see Eq. (\ref{Eq4}).
Phase jitter $\psi(t)$, if present, produces broadening of the $n$-th
component of the absorption line changing $\Gamma_{0}/2$ to $\Gamma
_{0}/2+n^{2}\left\langle (\delta\psi_{t})^{2}\right\rangle /2\tau_{c}$, where
$\left\langle (\delta\psi_{t})^{2}\right\rangle $ is the phase variance,
$\delta\psi_{t}=\psi(t)-\psi(0)$, and $\tau_{c}$ is a correlation time of the
phase change, see, for example, Refs. \cite{Shakhmuratov1,Shakhmuratov4}.
However, no difference in the linewidths of the components of the
M\"{o}ssbauer transmission spectrum are found in the experiments with grains.
One can find similar arguments in Ref. \cite{Monahan} where quantum beats of
recoil-free $\gamma$-radiation in time domain experiments are observed and
theoretically described. Interference terms $\left\langle J_{n}(M)J_{l-j}%
(M)\right\rangle $ producing quantum beats are not attenuated according to the
law $\exp[-j^{2}\sigma^{2}/2]$, which would be present if $\psi$ is normally
distributed about $\psi=0$ with variance $\sigma^{2}$. \begin{figure}[ptb]
\resizebox{0.9\textwidth}{!}{\includegraphics{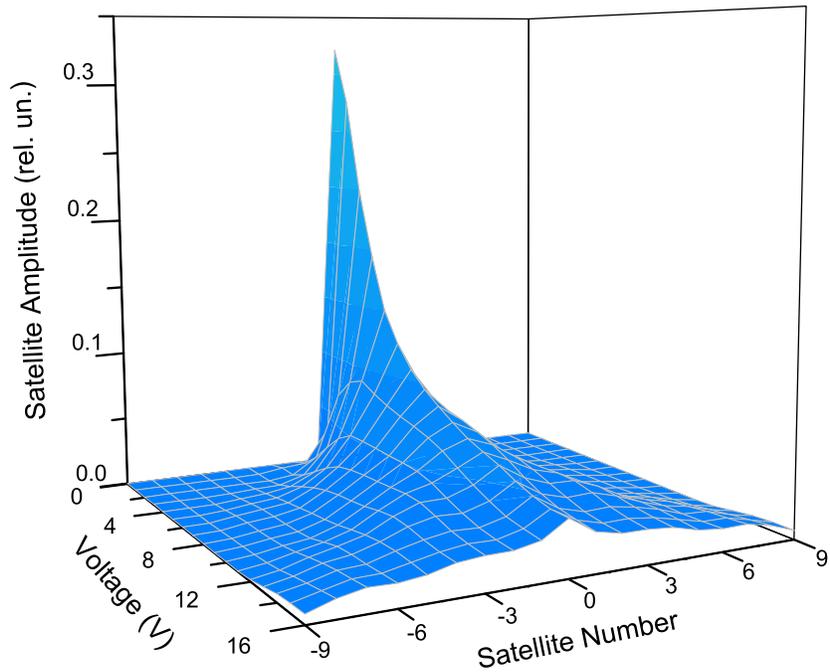}}\caption{(color on
line) Experimentally observed dependence of the absolute values of the
transmission dips, presented as the envelope line connecting the dips of the
spectral components, reverted upside down. Examples of the transmission
spectra are shown in Fig. 3. Front axis (x) shows the number of the spectral
component, vertical axis (z) is the absolute value of the depth of the
spectral component, which is $1-N_{SV\text{out}}(\omega_{A}-\omega_{S})$, and
y axis on the left is the voltage of the RF generator applied to the piezo
transducer.}%
\label{fig:2}%
\end{figure}

We model vibration of the powder pile by oscillating granular layers. A flat
layer of grains in the pile bottom is pushed up by the piezo transducer. Next
layer of grains is excited by the first layer, etc. We assume that vibration
amplitudes of the layers decrease from the pile bottom to the top. The
granular bed behaves much like a solid comoving with the vibrating piezo
transducer. However, because $\Lambda\gg1$ some grains may detach from the
transducer resulting in the particle convection. They move upward at the pile
center and downward at the edges. Vibration of grains can be decomposed into
coherent and incoherent parts. The coherent portion is dominant representing
the average of many measurements, while the incoherent portion gives less
contribution to M\"{o}ssbauer spectrum that diminishes when averaged in
repeated measurements \cite{Zhai}. This is because time scales of slow and
fast modes in grain dynamics are well separated. We also take into account
grain convection introducing the averaging over the angle between vertical
axis and direction along which an individual grain vibrates.

We model the M\"{o}ssbauer transmission spectrum by the expression%
\begin{equation}
N_{\text{out}}(\omega_{A}-\omega_{S})=e^{-\mu D}\left[  (1-f_{S}%
)+f_{S}\mathcal{L}(\omega_{A}-\omega_{S})\right]  , \label{Eq16}%
\end{equation}
where $\mu$ is nonresonant absorption coefficient, $D$ is the thickness of the
pile of powder, $f_{S}$ is the recoil-free fraction of $\gamma$-emission of
the source, and the function $\mathcal{L}(\omega_{A}-\omega_{S})$ is
\begin{equation}
\mathcal{L}(\omega_{A}-\omega_{S})=1-\sum_{n=-\infty}^{+\infty}C_{n}%
\mathcal{L}_{n}(\omega_{A}+n\Omega-\omega_{S}). \label{Eq17}%
\end{equation}
To simplify experimental data analysis, the function $\mathcal{L}_{n}%
(\omega_{A}+n\Omega-\omega_{S})$ is approximated by Lorentzian%
\begin{equation}
\mathcal{L}_{n}(\omega_{A}-\omega_{S})=\frac{B_{\exp}\Gamma_{\exp}^{2}%
}{(\omega_{A}+n\Omega-\omega_{S})^{2}+\Gamma_{\exp}^{2}}, \label{Eq18}%
\end{equation}
where $B_{\exp}$ together with $C_{n}$ define the maximum depth of the $n$-th
spectral component of the M\"{o}ssbauer transmission spectrum and
$\Gamma_{\exp}$ is the width of this component, which is taken the same for
all components. Parameters $B_{\exp}$ and $\Gamma_{\exp}$ are found from
fitting to the experimental spectra.

We assume that the vibration amplitude of the grains in the pile decays as
$a(x)=a_{0}\exp(-\gamma x)$, where $a_{0}$ is the amplitude at the pile
bottom, $x$ is the distance from the bottom to a particular upper layer of
grains, and $\gamma$ is a decay constant. Then, the coefficients $C_{n}$ can
be calculated according to Eq. (\ref{EqB34}), where $M(x)=M_{0}\exp(-\gamma
x)$ and $M_{0}$ is the modulation index at the pile bottom, see Appendix B.

In our model we assume that slow mode of granular dynamics producing particle
convection builds up chains of the strongly contacting grains, which form the
fastest ultrasound wave paths, see Ref. \cite{Zhai}. These paths are tortuous
slowly fluctuating in time. To take this slow process into account in the fast
mode of granular dynamics we introduce averaging over the angle $\alpha$
between the vertical axis and direction along which an individual particle
vibrates in the chain. Then, the coefficients $C_{n}$ in Eq. (\ref{Eq17}) can
be expressed as follows%
\begin{equation}
C_{n}=\frac{1}{D}\int_{0}^{D}dx\int_{0}^{\pi/2}d\alpha\frac{e^{-\alpha
^{2}/2\alpha_{m}^{2}}}{U(\alpha_{m})}J_{n}^{2}[M(x)\cos\alpha], \label{Eq19}%
\end{equation}
where%
\begin{equation}
U(\alpha_{m})=\alpha_{m}\sqrt{\frac{\pi}{2}}\operatorname{erf}\left(
\frac{\pi/2}{\sqrt{2}\alpha_{m}}\right)  , \label{Eq20}%
\end{equation}
and $\alpha_{m}^{2}$ is the variance of the angle distribution. It is also
possible introducing the dependence of $\alpha_{m}$ on distance $x$. However,
this modification complicates experimental data analysis and we decided to
take $\alpha_{m}$ as a constant parameter.

\section{Experimental results}

A conventional M\"{o}ssbauer spectrometer is used in experiment. The source,
$^{57}$Co:Rh, is mounted on the holder of the M\"{o}ssbauer drive Doppler
shifting the frequency of the radiation of the decaying nuclei. The absorber
was made of K$_{4}$Fe(CN)$_{6}\cdot3$H$_{2}$O powder with effective thickness
$T_{A}=13.2$. Actually, the absorber was a homogeneous mixture of powder with
natural abundance of $^{57}$Fe ($\sim2\%$) and enriched one (95$\%$). The
enriched powder was produced by M. N. Mikheev Institute of Metal Physics, Ural
Branch of Russian Academy of Sciences, Yekaterinburg. Average abundance of
$^{57}$Fe in powder mixture was nearly $50\%$. Physical thickness of the
absorber, $D$, was close to 100 $\mu$m. The preparation of the absorber is
described at the end of the Introduction in Sec. I . The source was placed
below the absorber and the detector was mounted above the absorber. This
vertical geometry of the experiment allowed to keep a powder on the surface of
the vibrating transducer due to gravity. As a transducer we used a
polyvinylidene fluoride (PVDF) piezo polymer film (thickness 28 $\mu$m, model
LDT0-28K, Measurement Specialties, Inc.). A piece of 10$\times$12 mm polar
PVDF film was coupled to a plexiglas backing of $\sim$2 mm thickness with
epoxy glue. The PVDF film transforms the sinusoidal signal from the
radio-frequency (RF) generator into a uniform vibration of the absorber nuclei
in the bottom of the powder pile, which is in a mechanical contact with the
PVDF film.

Examples of the experimentally observed spectra are shown in Fig. 3. They are
fitted by the theoretical model, Eq. (\ref{Eq16}). Three main fitting
parameters of the model are the modulation index $M_{0}$, the decay of the
vibration amplitude in the vertical direction, $\gamma$, and the standard
deviation, $\alpha_{m}$, of the particle vibration angle from vertical
direction. Optimal value of the first parameter for all voltages (the global
parameter) is $\gamma D=1.38\pm0.04$. Thus, the value of the vibration decay
constant is $\gamma=1.38\times10^{-4}$m$^{-1}$. The dependencies of the
parameters $M_{0}$ and $\alpha_{m}$ on the RF voltage V are shown in Fig. 4.
\begin{figure}[ptb]
\resizebox{1\textwidth}{!}{\includegraphics{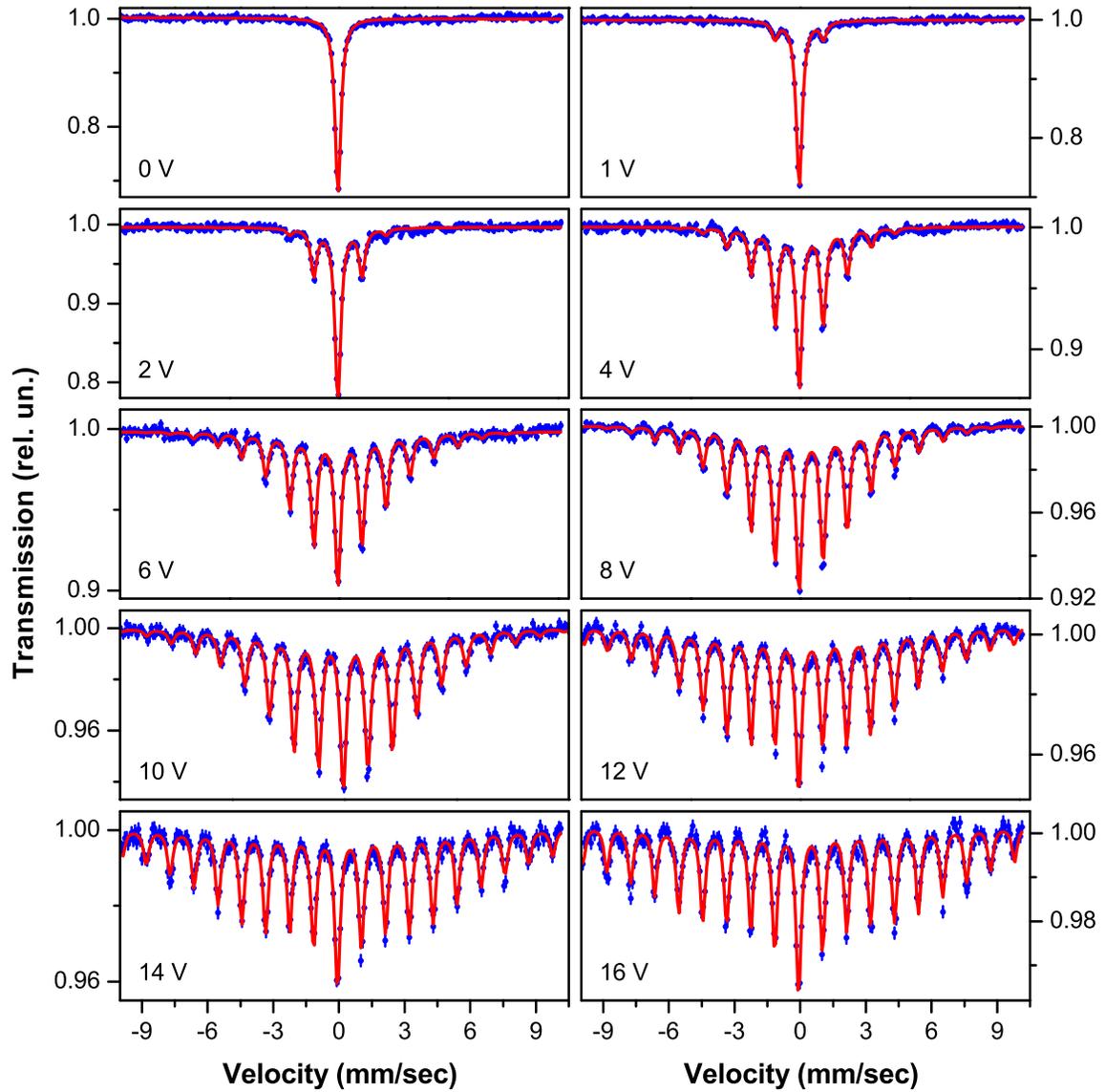}}\caption{(color on line)
The M\"{o}ssbauer transmission spectra for the vibrated powder. Horizontal
axis is the source velocity (in mm/sec) Doppler shifting the frequency of the
source. Vertical axis is the transmission probability, normalized to unity far
from resonance. The value of the RF voltage applied to the piezo transducer is
shown in the bottom of the left corner in each panel. Blue dots are
experimental data and red line is the theoretical fitting.}%
\label{fig:3}%
\end{figure}\begin{figure}[ptbptb]
\resizebox{0.8\textwidth}{!}{\includegraphics{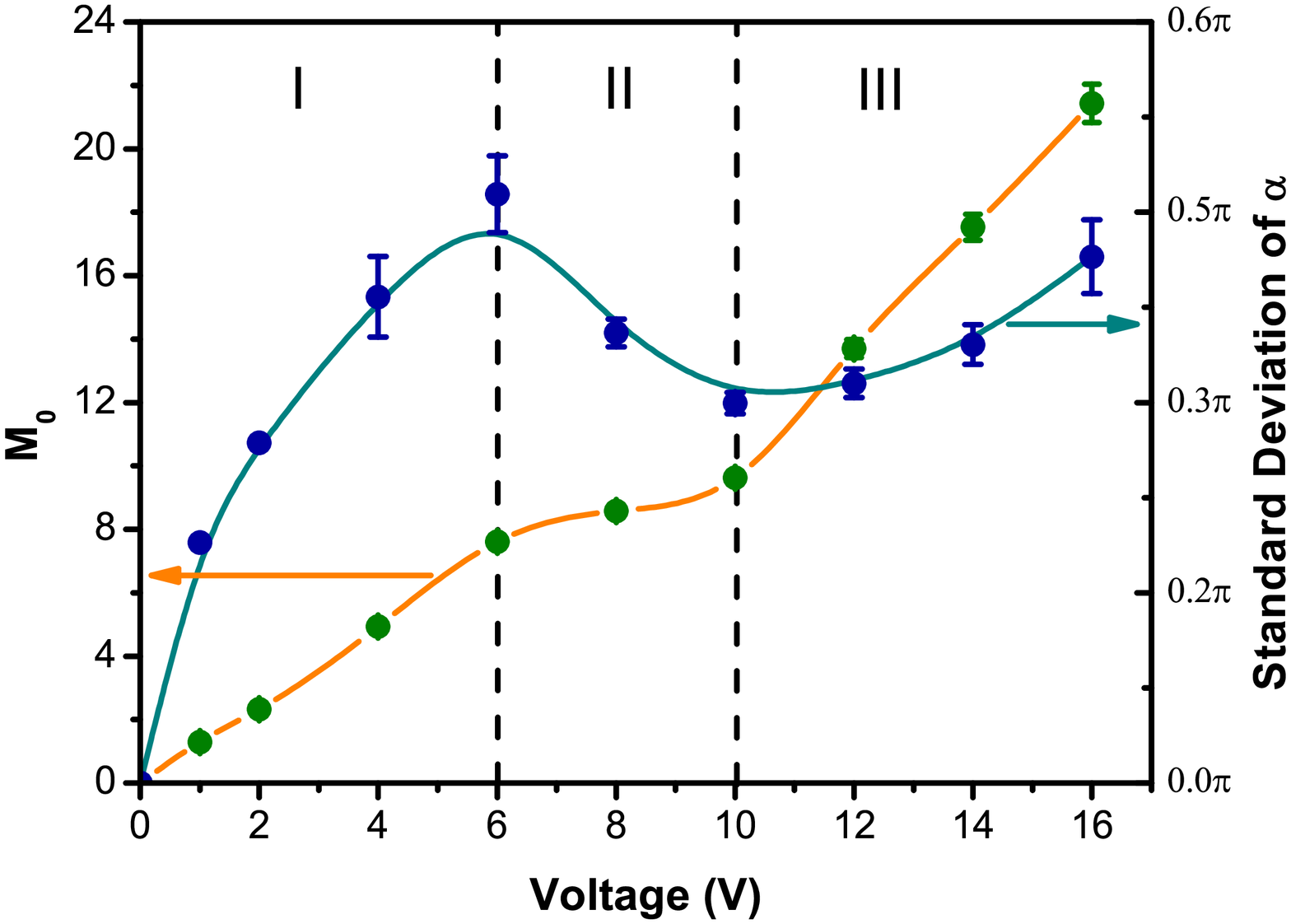}}\caption{(color on
line) Dependences of the modulation index $M_{0}$ for the powder bed (orange
line) and standard deviation of the vibration angle, $\alpha_{m}$, (dark cyan
line) on the RF voltage, V, applied to the piezo transducer. In both plots
dots correspond to the parameters derived from fitting and the lines
connecting the dots are shown for visualization. For some dots error bar
coincides or smaller than the size of the dot.}%
\label{fig:4}%
\end{figure}

Modulation index $M_{0}$, shown by orange line, monotonously rises with the RF
voltage increase. In the plot for $M_{0}$ versus V, there are three domains
with different slopes separated by vertical dashed lines. In the first domain
(I) standard deviation $\alpha_{m}$ of the angle $\alpha$, shown by dark cyan
line, gradually rises with the voltage increase reaching the value close to
$\alpha_{m}=\pi/2$. This means that at the right border of the first domain
appreciable fraction of grains vibrate in horizontal direction or do not
vibrate at all since the modulation index in Eq. (\ref{Eq19}) is
$M(x)\cos\alpha$. We may assume that when RF voltage is 6V many chains of the
strongly contacting grains become very tortuous or even broken. For this
voltage the parameter $M_{0}$ is 7, which gives vibration amplitude of the bed
grains $a=95.8$ pm.

As was mentioned in the Introduction, the kinetic energy of the vibrated
grain, $E_{\text{kin}}=ma^{2}\Omega^{2}/2$, plays an important role in
granular dynamics. If we introduce a potential energy of the grain lifted up
to the distance of one grain diameter $d$, i.e., $U_{g}=mgd$, where $m$ is the
mass of one grain, then we can define a parameter $W=E_{\text{kin}}/U_{g}$,
which can be expressed as $W=(a\Omega)^{2}/(2gd)$ or $W=(a/2d)\Lambda$. This
parameter is considered as dimensionless kinetic energy not depending on mass
of the grain, see, for example, Ref. \cite{Ristow}. Actually, the number $W$
represents the competition between the vibrational and gravitational energies,
see Ref. \cite{Bhateja}. At the right border of the first domain in Fig. 4 we
have $W=2.3$. This means that vibration energy is enough to lift up at least
two grains, one sitting on the top of the other, to the distance $d$.

Previous experiments with the stainless-steel foil demonstrated that polymer
piezo-transducer film vibrates in a uniform fashion without formation of
standing surface waves with nodes or traveling surface waves
\cite{Shakhmuratov2}. Therefore, we may expect that for small amplitudes $a$
when $W\ll1$ the powder after the preparation period forms densely packed
structure reminiscent polycrystal. When $W\sim1$, which is the fluidization
threshold, particles can move along the horizontal direction due to
collisions. In this region the horizontal dynamics are dominated by
particle-particle rather than particle-film interactions, see Ref.
\cite{Ristow,Olafsen}. This leads to the instabilities in the collective
motion of particles as energy is lost through interparticle collisions.
Nonuniform energy distribution arising from these instabilities in the
collective motion results in clustering in initially homogeneous granular
medium appearing as regions of high particle density and increased dissipation
rate \cite{Ristow}. We suppose that on the border of the domains I and II the
clusters are formed. Within the clusters particles come to rest in horizontal
direction \cite{Ristow}. Therefore, we see decrease of standard deviation
$\alpha_{m}$ in the domain II in Fig. 4.

The average size of particle clusters decreases with increasing acceleration
$\Lambda$ \cite{Ristow,Olafsen}. The second border between domains II and III
takes place at RF voltage 10V when $M_{0}=9$ and $a=1.23$ pm. For this voltage
we have $W=3.8$ and kinetic energy pumped to grains by transduced is enough to
lift up four particles to the distance $d$. Since $\alpha_{m}$ rises with
voltage increase in this domain, we suppose that the next phase transition in
granular dynamics takes place on the border between domains II and III.

\section{Conclusion}

We studied the transformation of M\"{o}ssbauer single parent line of the
vibrated powder absorber into a reduced intensity central line accompanied by
many sidebands due to the Raman scattering of the radiation field on the
vibrated nuclei. The intensities of the sidebands contain information about
the amplitudes of the mechanical vibrations of the grains and the amplitude
distribution along the propagation direction of $\gamma$-radiation. The
experimental spectra are fitted by the model containing decay of ultrasound
propagating from the bottom of the powder pile to its top. Also slow mode in
granular dynamics resulting in the grain convection is taken into account by
considering the chain of grains in hard contact, which form the fastest
ultrasound wave paths. We expect that our method will open a way for a new
kind of spectroscopical measurements of dynamical motion of granular material
subject to the vibrational excitation.

\section{Acknowledgements}

Experimental part of this work was partially funded by the Russian Foundation
for Basic Research (Grant No. 18-02-00845-a) and the Program of Competitive
Growth of Kazan Federal University, funded by the Russian Government. A. L. Z.
acknowledges support by the research grant of Kazan Federal University.

\section{Appendix A}

The propagation of $\gamma$-radiation through a resonant M\"{o}ssbauer medium
can be treated classically \cite{Lynch}. In this approach, the radiation
field, emitted by the source nucleus, after passing through a small diaphragm
is approximated as a plane wave propagating along the direction $\mathbf{z}$.
This field is described by
\begin{equation}
E_{S}(t-t_{0})=E_{0}\theta(t-t_{0})e^{-(i\omega_{S}+\Gamma_{0}/2)(t-t_{0}%
)+ikz}, \label{EqA1}%
\end{equation}
where $\Gamma_{0}$ is the inverse value of the lifetime of the excited state
of the emitting source nucleus. Here, for simplicity, the fraction of the
radiation field with recoil is disregarded and linewidth of the source nucleus
$\Gamma_{S}$ is assumed to be $\Gamma_{0}$.

The Fourier transform of the radiation field amplitude emitted by the source
nucleus is%
\begin{equation}
E_{S}(\omega)=\frac{E_{0}e^{ikz}}{\Gamma_{0}/2+i(\omega_{S}-\omega)}
\label{EqA2}%
\end{equation}
After passing through the absorber with a single resonance line this field is
transformed as (see Ref. \cite{Lynch})%
\begin{equation}
E_{\text{out}}(\omega)=E_{0}\frac{\exp\left[  ikz-\frac{T_{A}\Gamma_{A}%
/4}{\Gamma_{A}/2+i(\omega_{A}-\omega)}\right]  }{\Gamma_{0}/2+i(\omega
_{S}-\omega)}. \label{EqA3}%
\end{equation}
Here, again nonresonant absorption is disregarded. The linewidths $\Gamma_{S}$
and $\Gamma_{A}$ can be different from $\Gamma_{0}$ due to line broadening
mechanisms in the source and absorber, respectively.

The value%
\begin{equation}
N_{0}=\frac{\Gamma_{0}}{E_{0}^{2}}\int_{t_{0}}^{\infty}E_{S}(t-t_{0}%
)E_{S}^{\ast}(t-t_{0})dt, \label{EqA4}%
\end{equation}
which is proportional to the time integrated intensity of the radiation field
$I_{S}(t-t_{0})=\left\vert E_{S}(t-t_{0})\right\vert ^{2}$, can be considered
as a photon probability. For the emitted single photon this probability is
defined as equal to unity. The function%
\begin{equation}
N_{\text{out}}(\omega_{A}-\omega_{S})=\frac{\Gamma_{0}}{E_{0}^{2}}\int
_{0}^{\infty}E_{\text{out}}(\tau)E_{\text{out}}^{\ast}(\tau)dt \label{EqA5}%
\end{equation}
gives the probability of photon detection at the exit of the absorber. Far
from resonance ($\left\vert \omega_{A}-\omega_{S}\right\vert \gg\Gamma_{0}$)
this probability is unity. In resonance the detection probability of the
photon drops. Here, for simplicity, we take $\Gamma_{S}=\Gamma_{A}=\Gamma_{0}$.

According to Parseval's theorem we have
\begin{equation}
N_{\text{out}}(\omega_{A}-\omega_{S})=\frac{\Gamma_{0}}{2\pi E_{0}^{2}}%
\int_{-\infty}^{\infty}E_{\text{out}}(\omega)E_{\text{out}}^{\ast}%
(\omega)d\omega. \label{EqA6}%
\end{equation}
This expression gives the familiar in M\"{o}ssbauer spectroscopy formula for
$\gamma$-quanta absorption (see Ref. \cite{Gutlich})%
\begin{equation}
N_{\text{out}}(\omega_{A}-\omega_{S})=\frac{\Gamma_{0}}{2\pi}\int_{-\infty
}^{\infty}\frac{\exp\left[  -\frac{T_{A}(\Gamma_{0}/2)^{2}}{\left(  \Gamma
_{0}/2\right)  ^{2}+(\omega_{A}-\omega)^{2}}\right]  }{\left(  \Gamma
_{0}/2\right)  ^{2}+(\omega_{S}-\omega)^{2}}d\omega. \label{EqA7}%
\end{equation}
In exact resonance ($\omega_{A}=\omega_{S}$) the photon probability at the
exit of the absorber drops according to the expression (see Ref.
\cite{Vertes}) $N_{\text{out}}(0)=\exp(-T_{A}/2)I_{0}(T_{A}/2)$, where
$I_{0}(T_{A}/2)$ is the modified Bessel function of zero order.

\subsection{Vibrating absorber}

After passing through the absorber the source field $E_{AV}(\tau)$ in the
vibrated frame is transformed as
\begin{equation}
E_{AV\text{out}}(\tau)=\sum_{n=-\infty}^{+\infty}J_{n}(M)E_{n}(\tau
)e^{-i\omega_{S}\tau+in(\Omega\tau+\phi)}, \label{EqA8}%
\end{equation}
where%
\begin{equation}
E_{n}(\tau)e^{-i(\omega_{S}-n\Omega)\tau}=\frac{E_{0}}{2\pi}\int_{-\infty
}^{+\infty}\frac{\exp\left[  -i\omega\tau-\frac{T_{A}\Gamma_{A}/4}{\Gamma
_{A}/2+i(\omega_{A}-\omega)}\right]  }{\Gamma_{0}/2+i(\omega_{S}%
-n\Omega-\omega)}d\omega, \label{EqA9}%
\end{equation}
In the laboratory frame the radiation field is%
\begin{equation}
E_{L}(\tau)=e^{-iM\sin(\Omega\tau+\phi)}E_{AV\text{out}}(\tau), \label{EqA10}%
\end{equation}
which is polychromatic, i.e,
\begin{equation}
E_{L}(\tau)=\sum_{n=-\infty}^{+\infty}\sum_{m=-\infty}^{+\infty}J_{n}%
(M)J_{m}(M)E_{n}(\tau)e^{-i\omega_{S}\tau+i(n-m)(\Omega\tau+\phi)}.
\label{EqA11}%
\end{equation}
If all spectral components of the comb $E_{A\text{vib}}(\tau)$ in the
vibrating frame are far from resonance with the single line absorber, then
their amplitudes do not change, i.e., they are $E_{n}(\tau)J_{n}%
(M)=E_{in}(\tau)J_{n}(M)$, where%
\begin{equation}
E_{in}(\tau)=E_{0}\theta(\tau)e^{-\Gamma_{0}\tau/2}. \label{EqA12}%
\end{equation}
In this case the radiation field at the exit of the absorber, Eq.
(\ref{EqA11}), can be expressed in the lab frame as%
\begin{equation}
E_{L}(\tau)=e^{-i\omega_{S}\tau+i\varphi(\tau)-i\varphi(\tau)}E_{in}%
(\tau)=E_{S}(\tau), \label{EqA13}%
\end{equation}
where $\varphi(\tau)=M\sin(\Omega\tau+\phi)$. Therefore, the field does not
change and the probability of its registration is unity, i.e., equals to
$N_{0}=1$.

If $n$th component of the comb is close to resonance with the absorber and
other spectral components are far from resonance, then only the amplitude
$E_{n}(\tau)$ changes and we have%
\begin{equation}
E_{L}(\tau)=e^{-i\omega_{S}\tau-i\varphi(\tau)}\left\{  e^{i\varphi(\tau
)}E_{in}(\tau)-J_{n}(m)\left[  E_{in}(\tau)-E_{n}(\tau)\right]  e^{in(\Omega
\tau+\phi)}\right\}  . \label{EqA14}%
\end{equation}
This expression can be rewritten as follows%
\begin{equation}
E_{L}(\tau)=E_{S}(\tau)+E_{R}(\tau), \label{EqA15}%
\end{equation}
where $E_{S}(\tau)$ is the source field and%
\begin{equation}
E_{R}(\tau)=-e^{-i\omega_{S}\tau}\left[  E_{in}(\tau)-E_{n}(\tau)\right]
J_{n}(M)\sum_{m=-\infty}^{+\infty}J_{m}(M)e^{i(n-m)(\Omega\tau+\phi)}.
\label{EqA16}%
\end{equation}
is the field scattered by the vibrating absorber. The scattered field is
polychromatic containing the resonant for the absorber component with
frequency $\omega_{S}=\omega_{A}+n\Omega$ and Raman components $\omega
_{S}-(n-m)\Omega=$ $\omega_{A}+m\Omega$ with $m\neq n$. The resonantly
scattered component is in antiphase with the incident radiation reducing its
intensity due to destructive interference. The Raman components appear due to
inelastic scattering of the incident radiation field on the vibrating nuclei.
The amplitudes of these components depend on $J_{n}(m)$, i.e., on the
amplitude of the resonant component in the frequency comb $E_{AV}(\tau)$.

Fourier transform of the field $E_{R}(\tau)$ is%
\begin{equation}
E_{R}(\omega)=-E_{0}J_{n}(M)\sum_{m=-\infty}^{+\infty}\frac{J_{m}%
(M)e^{i(n-m)(\Omega t_{0}+\psi))}\left[  1-e^{-\frac{T_{A}\Gamma_{A}/4}%
{\Gamma_{A}/2+i(\omega_{A}+m\Omega-\omega)}}\right]  }{\Gamma_{0}%
/2+i[\omega_{S}-(n-m)\Omega-\omega]}. \label{EqA17}%
\end{equation}

\section{Appendix B}

Here we consider the propagation of $\gamma$-radiation through two vibrated
absorbers and four vibrated absorbers. These examples are given to derive a
generalized expression for the field transmitted through a pile of vibrated
grains. Vibration amplitudes of the test samples decrease along $\gamma
$-photon propagation direction step by step. From the results obtained for
these examples we may conclude what will be a result for the gradual change of
the vibration amplitude along a single absorber.

\subsection{Two vibrated absorbers}

Suppose we have two absorbers with the same resonance frequency $\omega_{A}$
and same effective thickness $T_{A}$. Both absorbers vibrate with frequency
$\Omega$. However the ampliteds and phases of vibrations are differen, which
are $a_{1}$, $\psi_{1}$ and $a_{2}$, $\psi_{2}$, respectively.

If $n$-th spectral component of the radiation field in the vibrating reference
frame rigidly bounded to the first absorber is close to resonance and other
components are far from resonance, then the radiation field at the exit of the
first absorber in the lab frame is%
\begin{equation}
E_{L1}(\tau)=E_{S}(\tau)+E_{1}(\tau)e^{-i\varphi_{1}(\tau)}, \label{EqB1}%
\end{equation}
where%
\begin{equation}
E_{1}(\tau)=-J_{n}(M_{1})E_{nSC}(\tau)e^{in(\Omega\tau+\phi_{1})},
\label{EqB2}%
\end{equation}
$\varphi_{1}(\tau)=M_{1}\sin(\Omega\tau+\phi_{1})$, $M_{1}=2\pi a_{1}/\lambda
$, $\phi_{1}=\Omega t_{0}+\psi_{1}$,%
\begin{equation}
E_{nSC}(\tau)=E_{S}(\tau)-E_{n}(\tau)e^{-i\omega_{S}\tau}, \label{EqB3}%
\end{equation}
$\ $and $E_{n}(\tau)$ is defined in Eq. (\ref{EqA9}).

In the reference frame of the second vibrated absorber the incident field is%
\begin{equation}
E_{2\text{in}}(\tau)=e^{i\varphi_{2}(\tau)}E_{L1}(\tau), \label{EqB4}%
\end{equation}
where $\varphi_{2}(\tau)=M_{2}\sin(\Omega\tau+\varphi_{2})$, and at the exit
of the absorber this field is transformed as%
\begin{equation}
E_{2\text{out}}(\tau)=E_{S}(\tau)e^{i\varphi_{2}(\tau)}+E_{11}(\tau
)e^{-i\varphi_{12}(\tau)}+E_{22}(\tau)+E_{12}(\tau), \label{EqB5}%
\end{equation}
where%
\begin{equation}
E_{11}(\tau)=-J_{n}(M_{1})E_{nSC}(\tau)e^{in(\Omega\tau+\phi_{1})},
\label{EqB6}%
\end{equation}%
\begin{equation}
E_{22}(\tau)=-J_{n}(M_{2})E_{nSC}(\tau)e^{in(\Omega\tau+\phi_{2})},
\label{EqB7}%
\end{equation}
are the fields produced due to scattering in the first and second absorbers,
respectively. These expressions contain the same function $E_{nSC}(\tau)$
since effective thickness of the absorbers is taken identical. The function%
\begin{equation}
E_{12}(\tau)=J_{n}(M_{1})J_{0}(M_{12})E_{nCSC}(\tau)e^{in(\Omega\tau+\phi
_{1})}, \label{EqB8}%
\end{equation}
describes consecutive scattering of the incident field in two absorbers, one
after another. The function $E_{nCSC}(\tau)$ is defined as%
\begin{equation}
E_{nCSC}(\tau)e^{in\Omega\tau}=\frac{E_{0}}{2\pi}\int_{-\infty}^{+\infty}%
\frac{\left[  1-e^{-\frac{T_{A}\Gamma_{A}/4}{\Gamma_{A}/2+i(\omega_{A}%
-\omega)}}\right]  ^{2}e^{-i\omega\tau}}{\Gamma_{0}/2+i(\omega_{S}%
-n\Omega-\omega)}d\omega. \label{EqB9}%
\end{equation}
The function $\varphi_{12}(\tau)=\varphi_{1}(\tau)-\varphi_{2}(\tau)$ in Eq.
(\ref{EqB5}) can be expressed as $\varphi_{12}(\tau)=M_{12}\sin(\Omega
\tau+\phi_{12})$, where $M_{12}=\sqrt{M_{1}^{2}+M_{2}^{2}-2M_{1}M_{2}\cos
(\phi_{1}-\phi_{2})}$ and $\phi_{12}=\tan^{-1}\left[  \frac{M_{2}\sin(\phi
_{1}-\phi_{2})}{M_{1}-M_{2}\cos(\phi_{1}-\phi_{2})}\right]  $.

In the laboratory frame the output field from the second absorber is%
\begin{equation}
E_{L2}(\tau)=E_{S}(\tau)+E_{11}(\tau)e^{-i\varphi_{1}(\tau)}+\left[
E_{22}(\tau)+E_{12}(\tau)\right]  e^{-i\varphi_{2}(\tau)}. \label{EqB10}%
\end{equation}
The spectral components of this field are%
\begin{equation}
E_{L2}(\tau)=\sum_{m=-\infty}^{+\infty}\mathcal{E}_{m}(\tau)e^{im\Omega\tau},
\label{EqB11}%
\end{equation}
where%
\begin{multline}
\mathcal{E}_{0}(\tau)=E_{S}(\tau)-E_{nSC}(\tau)\left[  J_{n}^{2}(M_{1}%
)+J_{n}^{2}(M_{2})\right]  +\\
+E_{nCSC}(\tau)J_{n}(M_{1})J_{0}(M_{12})J_{n}(M_{2})e^{in(\phi_{1}-\phi_{2})},
\label{EqB12}%
\end{multline}
and
\begin{multline}
\mathcal{E}_{m}(\tau)=-E_{nSC}(\tau)\left[  J_{n}(M_{1})J_{n-m}(M_{1}%
)e^{im\phi_{1}}+J_{n}(M_{2})J_{n-m}(M_{2})e^{im\phi_{2}}\right]  +\\
+E_{nCSC}(\tau)J_{n}(M_{1})J_{0}(M_{12})J_{n-m}(M_{2})e^{in(\phi_{1}-\phi
_{2})+im\phi_{2}}. \label{EqB13}%
\end{multline}
for $m\neq0$. We remind that here the case when $n$-th component of the comb
is in resonance with the single line absorber is considered.

Fourier transform of this field is
\begin{equation}
E_{L2}(\omega)=\sum_{m=-\infty}^{+\infty}\mathcal{E}_{m}(\omega-m\omega),
\label{EqB14}%
\end{equation}
where%
\begin{multline}
\mathcal{E}_{0}(\omega)=E_{S}(\omega)-E_{nSC}(\omega)\left[  J_{n}^{2}%
(M_{1})+J_{n}^{2}(M_{2})\right]  +\\
+E_{nCSC}(\omega)J_{n}(M_{1})J_{0}(M_{12})J_{n}(M_{2})e^{in(\phi_{1}-\phi
_{2})}, \label{EqB15}%
\end{multline}
for $m=0$, and%
\begin{multline}
\mathcal{E}_{m}(\omega-m\omega)=-E_{nSC}(\omega-m\Omega)\left[  J_{n}%
(M_{1})J_{n-m}(M_{1})e^{im\phi_{1}}+J_{n}(M_{2})J_{n-m}(M_{2})e^{im\phi_{2}%
}\right]  +\\
+E_{nCSC}(\omega-m\Omega)J_{n}(M_{1})J_{0}(M_{12})J_{n-m}(M_{2})e^{in(\phi
_{1}-\phi_{2})+im\phi_{2}}. \label{EqB16}%
\end{multline}
for $m\neq0$,%
\begin{equation}
E_{nSC}(\omega)=E_{0}\frac{\left[  1-e^{-\frac{T_{A}\Gamma_{A}/4}{\Gamma
_{A}/2+i(\omega_{A}+n\Omega-\omega)}}\right]  }{\Gamma_{0}/2+i(\omega
_{S}-\omega)}, \label{EqB17}%
\end{equation}
and%
\begin{equation}
E_{nCSC}(\omega)=E_{0}\frac{\left[  1-e^{-\frac{T_{A}\Gamma_{A}/4}{\Gamma
_{A}/2+i(\omega_{A}+n\Omega-\omega)}}\right]  ^{2}}{\Gamma_{0}/2+i(\omega
_{S}-\omega)}, \label{EqB18}%
\end{equation}
At exact resonance ($\omega_{S}=\omega_{A}+n\Omega$) the functions
$E_{nSC}(\omega)$, $E_{nCSC}(\omega)$ and $E_{nSC}(\omega-m\Omega)$,
$E_{nCSC}(\omega-m\Omega)$ have the same maxima%
\begin{equation}
E_{nSC}(\omega_{S})=E_{nSC}(\omega_{S}-m\Omega)=2E_{0}\left(  1-e^{-T_{A}%
/2}\right)  /\Gamma_{0}, \label{EqB19}%
\end{equation}%
\begin{equation}
E_{nCSC}(\omega_{S})=E_{nCSC}(\omega_{S}-m\Omega)=2E_{0}\left(  1-e^{-T_{A}%
/2}\right)  ^{2}/\Gamma_{0}, \label{EqB20}%
\end{equation}
however centered at differen frequencies, i.e., $E_{nSC}(\omega)$,
$E_{nCSC}(\omega)$ at $\omega=\omega_{S}$ and $E_{nSC}(\omega-m\Omega)$,
$E_{nCSC}(\omega-m\Omega)$ at $\omega=\omega_{S}-m\Omega$.

Finally we obtain that the detection probability of a single photon at the
exit of two vibrated absorbers is%
\begin{equation}
N_{\text{out}}(\omega_{A}-\omega_{S})=\frac{\Gamma_{0}}{2\pi E_{0}^{2}}%
\sum_{m=-\infty}^{+\infty}\int_{-\infty}^{\infty}\left\vert \mathcal{E}%
_{m}(\omega-m\omega)\right\vert ^{2}d\omega. \label{EqB21}%
\end{equation}

We make further simplifications in our model assuming that two absorbers
vibrate with the same frequency and their amplitudes satisfy inequality
$a_{1}>a_{2}$. If vibration phases of two samples are the same, $\psi_{1}%
=\psi_{2}$, then in a time period when transducer pushes the first absorber
up, the second absorber moves also up but with smaller amplitude. Therefore,
in the vibrating frame of the first absorber the second absorber moves towards
the first sample and we have compression of powder in these samples. In the
second time period when transducer moves down we have decompression since the
second sample delays with respect to the first sample. In the case of equal
phases we have $M_{12}=M_{1}-M_{2}$. If the difference between the amplitudes
$a_{1}$ and $a_{2}$ is small, the modulation index $M_{12}$ in Eqs.
(\ref{EqB15}-\ref{EqB16}) is also small.

\subsection{Four vibrated absorbers}

We calculated transmission of a single $\gamma$-photon through the four
vibrating absorbers. For simplification of the result we take the same phases
$\psi$ of their vibrations and impose a condition $a_{1}>a_{2}>a_{3}>a_{4}$,
where $a_{i}$ is the vibration amplitude of the $i$th sample. The result is%
\begin{equation}
E_{L4}(\omega)=\sum_{m=-\infty}^{+\infty}\mathcal{E}_{m}(\omega-m\omega
)e^{im\phi}, \label{EqB22}%
\end{equation}%
\begin{multline}
\mathcal{E}_{0}(\omega)=E_{S}(\omega)-E_{nSC1}(\omega)\sum_{i=1}^{4}J_{n}%
^{2}(M_{i})+\\
+E_{nSC2}(\omega)\sum_{\substack{i,l\\(i>l,i\neq4)}}^{4}J_{n}(M_{i}%
)J_{0}(M_{il})J_{n}(M_{l})-\\
-E_{nSC3}(\omega)\sum_{\substack{i,l,p\\(i>l>p,i\neq4)}}^{4}J_{n}(M_{i}%
)J_{0}(M_{il})J_{0}(M_{lp})J_{n}(M_{p})+\\
+E_{nSC4}(\omega)J_{n}(M_{1})J_{0}(M_{12})J_{0}(M_{23})J_{0}(M_{34}%
)J_{n}(M_{4}), \label{Eq23}%
\end{multline}%
\begin{multline}
\mathcal{E}_{m}(\omega-m\Omega)=-E_{nSC1}(\omega-m\Omega)\sum_{i=1}^{4}%
J_{n}(M_{i})J_{n-m}(M_{i})+\\
+E_{nSC2}(\omega-m\Omega)\sum_{\substack{i,l\\(i>l,i\neq4)}}^{4}J_{n}%
(M_{i})J_{0}(M_{il})J_{n-m}(M_{l})+\\
-E_{nSC3}(\omega-m\Omega)\sum_{\substack{i,l,p\\(i>l>p,i\neq4)}}^{4}%
J_{n}(M_{i})J_{0}(M_{il})J_{0}(M_{lp})J_{n-m}(M_{p})+\\
+E_{nSC4}(\omega-m\Omega)J_{n}(M_{1})J_{0}(M_{12})J_{0}(M_{23})J_{0}%
(M_{34})J_{n-m}(M_{4}), \label{Eq24}%
\end{multline}
where $M_{i}$ is the modulation index for $i$th sample and $M_{il}=M_{i}%
-M_{l}$. Summation over positive integers $i,l$ contains combinations 1,2;
1,3; 1,4; 2,3; 2,4; 3,4, and summation over $i,l,p$ contains 1,2,3; 1,2,4;
1,3,4; 2,3,4. The functions $E_{nSCl}(\omega)$ are%
\begin{equation}
E_{nSCk}(\omega)=E_{0}\frac{\left[  1-e^{-\frac{T_{A}\Gamma_{A}/4}{\Gamma
_{A}/2+i(\omega_{A}+n\Omega-\omega)}}\right]  ^{k}}{\Gamma_{0}/2+i(\omega
_{S}-\omega)}. \label{EqB25}%
\end{equation}
If all four absorbers vibrate with the same amplitude $a$ and modulation index
is $M=2\pi a/\lambda$, then the spectral components of the output field are%
\begin{equation}
\mathcal{E}_{0}(\omega)=E_{S}(\omega)\left\{  \left[  1-J_{n}^{2}(M)\right]
+J_{n}^{2}(M)\sum_{k=0}^{4}\binom{4}{k}(-1)^{k}T^{k}(\omega)\right\}  ,
\label{EqB26}%
\end{equation}%
\begin{equation}
\mathcal{E}_{m}(\omega-m\Omega)=-E_{S}(\omega-m\Omega)J_{n}(M)J_{n-m}%
(M)\left[  1-\sum_{k=0}^{4}\binom{4}{k}(-1)^{k}T^{k}(\omega-m\Omega)\right]  ,
\label{EqB27}%
\end{equation}
where $\binom{4}{k}$ is the binomial coefficient and%
\begin{equation}
T^{k}(\omega)=\left[  1-e^{-\frac{T_{A}\Gamma_{A}/4}{\Gamma_{A}/2+i(\omega
_{A}+n\Omega-\omega)}}\right]  ^{k}. \label{EqB28}%
\end{equation}
Since the summation gives%
\begin{equation}
\sum_{k=0}^{4}\binom{4}{k}(-1)^{k}T^{k}(\omega)=\left[  1-T^{k}(\omega
)\right]  ^{4}=e^{-\frac{T_{A}\Gamma_{A}}{\Gamma_{A}/2+i(\omega_{A}%
+n\Omega-\omega)}}, \label{EqB29}%
\end{equation}
we obtain for this case the same expression as for the single vibrated
absorber, Eqs. (\ref{Eq12}),(\ref{Eq13}), but with the effective thickness
$4T_{A}$, where factor $4$ comes from the number of the absorbers. Similar
result is valid for the absorber consisting of infinite number of layers
vibrated with the same amplitude.

In the case of different vibration amplitudes we numerically found that the
dominant contribution to the output field is given by the term
\begin{equation}
\mathcal{E}_{0}(\omega)=E_{S}(\omega)\left[  1-T(\omega)\sum_{i=1}^{4}%
J_{n}^{2}(M_{i})\right]  , \label{EqB30}%
\end{equation}
and the detection probability of a single photon at the exit of the vibrated
absorbers is approximated as%
\begin{equation}
N_{\text{out}}(\omega_{A}-\omega_{S})=1-\sum_{n=-\infty}^{+\infty}\sum
_{i=1}^{4}J_{n}^{2}(M_{i})F(\omega_{A}+n\Omega-\omega_{S}), \label{EqB31}%
\end{equation}
where $F(\omega_{A}+n\Omega-\omega_{S})$ is the spectral function, which is
zero far from resonance and has a minimum each time when $\omega_{S}%
=\omega_{A}+n\Omega$.

Qualitative arguments validating this approximation are the following.
Physical thickness of the pile of powder is about 100 $\mu$m while the median
size of grains is 1.3 $\mu$m. The grains are in physical contact with each
other and we have nearly 100 grains contacting in vertical direction with the
particle in the pile bottom sitting on the transducer. Due to friction the
vibration amplitude of grains decreases from the pile bottom to the top. If we
take diameter of a single particle as $d$, then the transmission function of
$\gamma$-photon for each grain is%
\begin{equation}
T(\omega)=\sum_{k=1}^{\infty}\frac{d^{k}}{2k!}\left[  \frac{\alpha_{A}%
\Gamma_{A}/2}{\Gamma_{A}/2+i(\omega_{A}+n\Omega-\omega)}\right]  ^{k},
\label{EqB32}%
\end{equation}
where $\alpha_{A}$ is the absorption coefficient of the absorbing material,
which gives effective thickness of a single grain $T_{AG}=\alpha_{A}d$. If
$T_{AG}\ll1$, we can keep in this expression only the first term proportional
to $d$. Then, the power spectrum of the output field is approximated as%
\begin{equation}
\left\vert \mathcal{E}_{0}(\omega)\right\vert ^{2}=\left\vert E_{S}%
(\omega)\right\vert ^{2}-2\left\vert E_{S}(\omega)E_{S}^{\ast}(\omega)T^{\ast
}(\omega)\right\vert \sum_{i=1}^{100}J_{n}^{2}(M_{i}), \label{EqB33}%
\end{equation}
where the terms proportional to $d^{2}$ are disregarded. The summation can be
replaced by integral, which gives%
\begin{equation}
N_{\text{out}}(\omega_{A}-\omega_{S})=1-\sum_{n=-\infty}^{+\infty}f(\omega
_{A}+n\Omega-\omega_{S})\int_{0}^{Nd}J_{n}^{2}[M(x)]dx, \label{EqB34}%
\end{equation}
where $N$ is the number of particles in the vertical chain of grains and the
function $f(\omega_{A}+n\Omega-\omega_{S})$ is the normalized function
$F(\omega_{A}+n\Omega-\omega_{S})$ to make $N_{\text{out}}(\omega_{A}%
-\omega_{S})$ dimensionless.


\begin{thebibliography}{99}                                                                                               %


\bibitem {Duran}J. Duran, \textit{Sands, powders, and grains: an introduction
to the physics of granular materials}, Part of the \textit{Partially ordered
system} book series (Editorial Board: L. Lam and D. Langevin, Springer
Science+Business Media, New York 2000) P.1-214. DOI 10.1007/978-1-4612-0499-2

\bibitem {Jaeger}H. M. Jaeger and S. R. Nagel, Physics of the Granular State,
Science \textbf{255}, 1524 (1992).

\bibitem {Ristow}G. H. Ristow, \textit{Pattern formation in granular
materials}, Part of the \textit{Springer Tracts in Modern Physics} book
series, Vol. 164 (Ed. G. H\"{o}hler, Springer-Verlag, Berlin Heidelberg 2000)
P. 1-161. DOI 10.1007/BFb0110577

\bibitem {Eshuis}P. Eshuis, K. van der Weele, D. van der Meer, R. Bos, and D.
Lohse, Phase diagram of vertically shaken granular matter, Physics of Fluids
\textbf{19}, 123301 (2007).

\bibitem {Herrmann}Physics of Dry Granular Media, edited by H. J. Herrmann,
J.-P. Hovi, and S. Luding, NATO ASI Series, Series E: Applied Sciences - Vol.
350 (Springer-Science+Business Media, Dordrecht 1998) P. 1-711. DOI 10.1007/978-94-017-2653-5

\bibitem {Klongboonjit}S. Klongboonjit and C. S. Campbell, Convection in deep
vertically shaken particle beds. I. General features, Physiscs of Fluids
\textbf{20}, 103301 (2008).

\bibitem {Liu}C.-h. Liu, S. R. Nagel, D. A. Schecter, S. N. Coppersmith, S.
Majumdar, O. Narayan, T. A. Witten, Force Fluctuation in Bead Packs, Science
\textbf{269}, 513 (1995).

\bibitem {Nagel}H. M. Jaeger, S. R. Nagel, and R. P. Berhinger, Granular
solids, liquids, and gases, Reviews of Modern Physics \textbf{68}, 1259 (1996).

\bibitem {Lu}X. Lu, S. Yang, and J. R. G. Evans, Ultrasound-assisted
microfeeding of fine powders, Particuology \textbf{6}, 2 (2008).

\bibitem {Majmudar}T. S. Majmudar and R. P. Behringer, Contact force
measurements and stress-induced anisotropy in granular materials, Nature
\textbf{435}, 1079 (2005).

\bibitem {Clement}E. Cl\'{e}ment, J. Duran, and J. Rajchenbach, Experimental
study of heaping in a two-dimntional "sandpile", Phys. Rev. Lett. \textbf{69},
1189 (1992).

\bibitem {Nagel2}E. E. Ehrichs, H. M. Jaeger, G. S. Karczmar, J. B. Knight, V.
Yu. Kuperman, and S. R. Nagel, Granular convection observed by magnetic
resonance imaging, Science \textbf{267}, 1632 (1995).

\bibitem {Bougie}J. Bougie, S. J. Moon, J. B. Swift, and H. L. Swinney, Shocks
in vertically oscillating granular layers, Phys. Rev E \textbf{66}, 051301 (2002).

\bibitem {Harada}S. Harada, S. Takagi, and Y. Matsumoto, Wave propagation in a
dynamic system of soft granular materials, Phys. Rev E \textbf{67}, 061305 (2003).

\bibitem {Amirifar}R. Amirifar, K. Dong, Q. Zeng, and X. An, Bimodal
self-assembly of granular spheres under vertical vibration, Soft Matter
\textbf{15}, 5933 (2019).

\bibitem {Harwood}C. F. Harwood, Powder segragation due to vibration, Powder
Technology \textbf{16}, 51 (1977).

\bibitem {Buck}M. Buckingham, Theory of acoustic attenuation, dispersion, and
pulse propagation in unconsolidated granular materials including marine
sediments, J. Acous. Soc. Am. \textbf{102}, No. 5, 2579 (1997).

\bibitem {Zhai}C. Zhai, E. B. Herbold, and R. C. Hurley, The influence of
packing structure and interparticle forces on ultrasound transmission in
granular media, PNAS \textbf{117}, 16234 (2020).

\bibitem {Mehta}A. Mehta, \textit{Granular physics} (Cambridge University
Press, New York 2007).

\bibitem {Shakhmuratov1}R. N. Shakhmuratov, F. G. Vagizov, Application of the
M\"{o}ssbauer effect to the study of subnanometer harmonic displacements in
thin solids, Phys. Rev. B \textbf{95}, 245429 (2017).

\bibitem {Shakhmuratov2}R. N. Shakhmuratov, F. G. Vagizov, M\"{o}ssbauer
method for measuring subangstrom displacements in thin films, JETP Letters
\textbf{108}, 772 (2018).

\bibitem {Cranshaw}T. E. Cranshaw and P. Reivari, A M\"{o}ssbauer study of the
hyperfine spectrum of $^{57}$Fe, using ultrasonic calibration, Proc. Phys.
Soc. \textbf{90}, 1059 (1967).

\bibitem {Mishroy}J. Mishroy and D. I. Bolef, Interaction of ultrasound with
M\"{o}ssbauer gamma rays, in \textit{M\"{o}ssbauer Effect Methodology}, edited
by I. J. Gruverman (Plenum Press, Inc. New York, 1968), Vol. 4, P. 13-35.

\bibitem {Asher}J. Asher, T. E. Cranshow, and D. A. O'Conner, The observation
of sidebands produced when monochromatic radiation passes through a vibrated
resonant medium, J. Phys. A: Math., Nucl. Gen., \textbf{7}, 410 (1974).

\bibitem {Tsankov}L. T. Tsankov, Experimental observations on the resonant
amplitude modulation of M\"{o}ssbauer gamma rays, J. Phys. A: Math. Gen.
\textbf{14}, 275 (1981).

\bibitem {Shvydko}Yu. V. Shvyd'ko and G. V. Smirnov, Enhanced yield into the
radiative channel in Raman nuclear resonant forward scattering, J. Phys.:
Condens. Matter \textbf{4}, 2663 (1992).

\bibitem {Shakhmuratov3}R. N. Shakhmuratov, F. G. Vagizov, M\"{o}ssbauer
method for studying vibrations in a granular medium excited by ultrasound,
JETP Letters \textbf{111}, 167 (2020).

\bibitem {Gutlich}P. G\"{u}tlich, E. Bill, and A. X. Trautwein,
\textit{M\"{o}ssbauer Spectroscopy and Transition Metal Cemistry: Fundamentals
and Applications} (Springer-Verlag, Berlin, Heidelberg, 2011).

\bibitem {Mkrtchyan77}A. R. Mkrtchyan, A. R. Arakelyan, G. A. Arutyunyan, and
L. A. Kocharyan, Oscillations of the M\"{o}ssbauer spectrum line intensity
following modulation by coherent ultrasound, Pis'ma Zh. Eksp. Teor. Fiz. Pisma
\textbf{26}, 599 (1977) [JETP Lett. \textbf{26}, 449 (1977)].

\bibitem {Mkrtchyan79}A. R. Mkrtchyan, G. A. Arutyunyan, A. R. Arakelyan, and
R. G. Gabrielyan, Modulation of M\"{o}ssbauer radiation by coherent ultrasounic
excitation in crystals, Phys. Stat. Sol. b \textbf{92}, 23 (1979).

\bibitem {Radeon}Y. V. Radeonychev, I. R. Khairullin, F. G. Vagizov, M.
Scully, and O. Kocharovskaya, Observation of Acoustically Induced Transparency
for $\gamma$-radiation, Phys. Rev. Lett. \textbf{124}, 163602 (2020).

\bibitem {Shakhmuratov4}R. N. Shakhmuratov, A. Szabo, Phase-noise influence on
coherent transients and hole burning, Phys. Rev. A \textbf{58}, 3099 (1998).

\bibitem {Monahan}J. E. Monahan, and G. J. Perlow, Theoretical description of
quantum beats of recoil-free $\gamma$ radiation, Phys. Rev. A \textbf{20},
1499 (1979).

\bibitem {Bhateja}A. Bhateja, I. Sharma, and J. K. Singh, Scaling of granular
temperature in vibro-fluidized grains, Physics of Fluids \textbf{28}, 043301 (2016).

\bibitem {Olafsen}J. S. Olafsen, J. S. Urbach, Clustering, order, and collapse
in a driven granular monolayer, Phys. Rev. Lett. \textbf{81}, 4369 (1998).

\bibitem {Lynch}F. J. Lynch, R. E. Holland, and M. Hamermesh, Time dependence
of resonantly filtered gamma rays form $^{57}$Fe, Phys. Rev. \textbf{120}, 513 (1960).

\bibitem {Vertes}A. V\'{e}rtes, L. Korez, and K. Burger, \textit{M\"{o}ssbauer
Spectroscopy} (Elsevier, New York, 1979).
\end{thebibliography}
\end{document}